\documentclass[twocolumn,prb,showpacs,multicol,amsmath,amssymb]{revtex4-1}
\usepackage{graphicx}
\usepackage{epstopdf}
\usepackage{subfigure}

\newcommand{\be}{\begin{equation}}
\newcommand{\ee}{\end{equation}}

\newcommand{\bea}{\begin{eqnarray}}
\newcommand{\eea}{\end{eqnarray}}
\newcommand{\bd}{\begin{displaymath}}
\newcommand{\ed}{\end{displaymath}}
\newcommand{\ba}{\begin{array}}
\newcommand{\ea}{\end{array}}
\newcommand{\bi}{\begin{itemize}}
\newcommand{\ei}{\end{itemize}}
\newcommand{\bc}{\begin{center}}
\newcommand{\ec}{\end{center}}
\newcommand{\bfl}{\begin{flushleft}}
\newcommand{\efl}{\end{flushleft}}
\newcommand{\bfr}{\begin{flushright}}
\newcommand{\efr}{\end{flushright}}

\newcommand{\bQ}{{\bf Q}}

\newcommand{\bk}{{\bf k}}

\def\ua{\uparrow}
\def\da{\downarrow}
\def\ket#1{\left\vert #1 \right\rangle}

\def\bk{{\bf k}}  
\def\bQ{{\bf Q}}
\def\Ce{CeCoIn$_5$}
\def\dwave{$d_{x^2-y^2}$-wave~}
\def\da{\downarrow} \def\ua{\uparrow} 

\def\6{\partial}

\def\bra{\langle}
\def\ket{\rangle}
\def\={\!\!\!&=&\!\!\!}
\def\+{\!\!\!&&\!\!\!+~}
\def\-{\!\!\!&&\!\!\!-~}

\begin{document}
\date{\today}
\title{Field induced spin exciton doublet splitting in 
$d_{x^2-y^2}$-wave 115-heavy electron superconductors}

\author{Alireza \surname{Akbari}}
\email{akbari@cpfs.mpg.de}

\author{Peter \surname{Thalmeier}}

 \affiliation{
Max Planck Institute for the  Chemical Physics of Solids, D-01187 Dresden, Germany
 }

 \begin{abstract}
 We investigate the spin-exciton modes in the superconducting $d_{x^2-y^2}$ state of  CeMIn$_5$  heavy fermion compounds
 found at the antiferromagnetic wave vector by inelastic neutron scattering.
We present a theoretical model that explains the field dependence for both field directions. 
 We show that the recently observed splitting of the spin exciton doublet in  CeCoIn$_5$ into two non-degenerate  modes for in-plane field appears naturally in this model. This is due to the spin anisotropy of g- factors and quasiparticle interactions which lead to different resonant conditions for the dynamic susceptibility components. We predict that the splitting of the spin resonance doublet becomes strongly nonlinear for larger fields when the energy of both split components decreases. For field along the tetragonal axis no splitting but only a broadening of the resonance is found in agreement with experiment.
 \end{abstract}


\maketitle

%%%%%%%%%%%%%%%%%%%%%%%%%%%%%%%%%%%%%%%%%%%%%%%%%%%%%%%%%%%%%%%%%%%%%%%%%%
%%%%%%%%%%%%%%%%%%%%%%%%%%     Section I      %%%%%%%%%%%%%%%%%%%%%%%%%%%%
%%%%%%%%%%%%%%%%%%%%%%%%%%%%%%%%%%%%%%%%%%%%%%%%%%%%%%%%%%%%%%%%%%%%%%%%%%
\section{Introduction}

In unconventional superconductors quasiparticle excitations that determine low temperature thermodynamics and response exhibit a generally anisotropic gap $\Delta(\bk)$ with posible node lines on the Fermi surface (FS). In addition to these single particle excitations collective excitations may appear. The collective oscillation of the superfluid density or 'Higgs mode' that belongs to the singlet spin sector is diffcult to observe. In addition collective spin triplet excitations may be present below the gap edge that are formed as bound states due to quasiparticle interactions. They are accessible directly by inelastic neutron scattering (INS) and have been found in a considerable number in unconventional superconductors. In heavy fermion compounds their typical energy is in the range of just one meV. The most clearcut example belongs to the class of 115 superconductors, CeMIn$_5$  (M= Rh, Ir and Co). They have attracted great interest because of coexisting and competing antiferromagnetic (AF) and superconducting (SC) states\cite{Thompson:01,Zapf:01,Sarrao:07,Thompson:12}, and in particular due to the possible existence of a Fulde-Ferrell-Larkin-Ovchinnikov (FFLO) phase in \Ce \cite{Bianchi:03}.\\

The highest superconducting critical temperature  of this  family appears in  \Ce  ~with T$_c= 2.3$K, and the question of the gap symmetry in these compounds has been intensely discussed \cite{Petrovic:01,Izawa:01, An:10}.
A powerful indirect method to study the unconventional gap symmetry is provided by the spin resonance peak which may appear in inelastic neutron scattering (INS). Such a pronounced spin resonance has been observed in CeCoIn$_5$ and La substituted crystals  at $\omega_r/2\Delta_1 = 0.65$ in the superconducting state by INS where $2\Delta_1$ is the main quasiparticle gap obtained from tunneling experiments \cite{Rourke:2005}. It is confined to a narrow region around the AF wave vector $\bQ=(\frac{1}{2},\frac{1}{2},\frac{1}{2})$  \cite{Stock:08}.  A theoretical calculation  of the dynamical spin response using realistic Fermi surface \cite{Eremin:08} shows that the resonance can appear only for the \dwave gap symmetry but not for $d_{xy}$- type gap because the precondition $\Delta(\bk+\bQ)=-\Delta(\bk)$ for a spin resonance \cite{Bulut:1996} is only fulfilled in the former case. Indeed the $d_{x^2-y^2}$ gap symmetry has also been found by thermal conductivity \cite{Izawa:01} and specific heat measurements \cite{An:10} in rotating magnetic fields.

The existence of a spin exciton resonance in the SC phase is a well established many-body effect which has also been observed in  other unconventional superconductors like high-Tc cuprates\cite{Mignod:91}, heavy fermion metals UPd$_2$Al$_3$\cite{Sato:01, Metoki:97, Chang:07} and CeCu$_2$Si$_2$\cite{Stockert:08,Stockert:11},  and in particular  in many  Fe-pnictide compounds \cite{Christianson:08}. The appearance of the resonance  depends sensitively on the type of unconventional Cooper-pairing and provides a powerful criterion to eliminate certain forms of pairing when a resonance is observed. These interpretations, however, all depend considerably on theoretical phenomenological model features like FS nesting properties, nodal positions and  momentum dependence of the gap as well as the size and anisotropy of quasiparticle interactions.
 
The spin exciton is a triplet excitation in the isotropic case since it should appear as a pole or resonance  in all three components of the susceptibility tensor. In principle,  in presence of the magnetic field it should  split into three modes with different polarization (left and right handed as well as longitudinal). The field-induced splitting of spin excitons  has been predicted for the cuprates\cite{Ismer:07} but was sofar never identified experimentally in any unconventional superconductor.

Recently an apparent field induced spin-exciton splitting has been  found for the first time in \Ce~ by INS and has given rise to an interesting debate. In Ref.~\onlinecite{Stock:12} a splitting into two (rather than the expected three) modes was found for field in the tetragonal plane and no splitting for perpendicular field. On the other hand no splitting was reported in Refs.~(\onlinecite{Panarin:11a, Panarin:11b}) for both field directions and  only  an increased broadening of the peak is observed \cite{Panarin:11a}. Furthermore it has been proposed that the FFLO-type inhomogeneous superconducting `{\bf Q}-phase' \cite{Yanase:09,Aperis:2010,Kenzelmann:10,Kumagai:11} is due to the condensation of the lower split-off branch of spin excitons close to the upper critical field \cite{Stock:12,Michal:11}.

In this paper we present a theoretical analysis to address the field splitting of the spin resonance which have been observed recently in INS experiments \cite{Stock:12} for the first time in an unconventional superconductor. We investigate whether the conventional picture of this mode as a triplet bound state of quasiparticles can explain these observations. We clarify the appropriate conditions for the splitting to occur for the different field directions as well as the number and  field dependence of the split spin exciton modes.

%%%%%%%%%%%%%%%%%%%%%%%%%%%%%%%%%%%%%%%%%%%%%%%%%%%%%%%%%%%%%%%%%%%%%%%%%%
%%%%%%%%%%%%%%%%%%%%%%%%%%     Section II      %%%%%%%%%%%%%%%%%%%%%%%%%%%%
%%%%%%%%%%%%%%%%%%%%%%%%%%%%%%%%%%%%%%%%%%%%%%%%%%%%%%%%%%%%%%%%%%%%%%%%%%

\section{Theoretical Model}
\label{secII}
%%%%%%%%% %%%%%%%%%%%%%%%%%%%%%% figure %%%%%%%%%%%%%%%%%%%%%%%%%%%%%%%%%%%%%%%%%
 \begin{figure}
    \centering
        \subfigure[]
    {
        \includegraphics[width=0.46\linewidth]{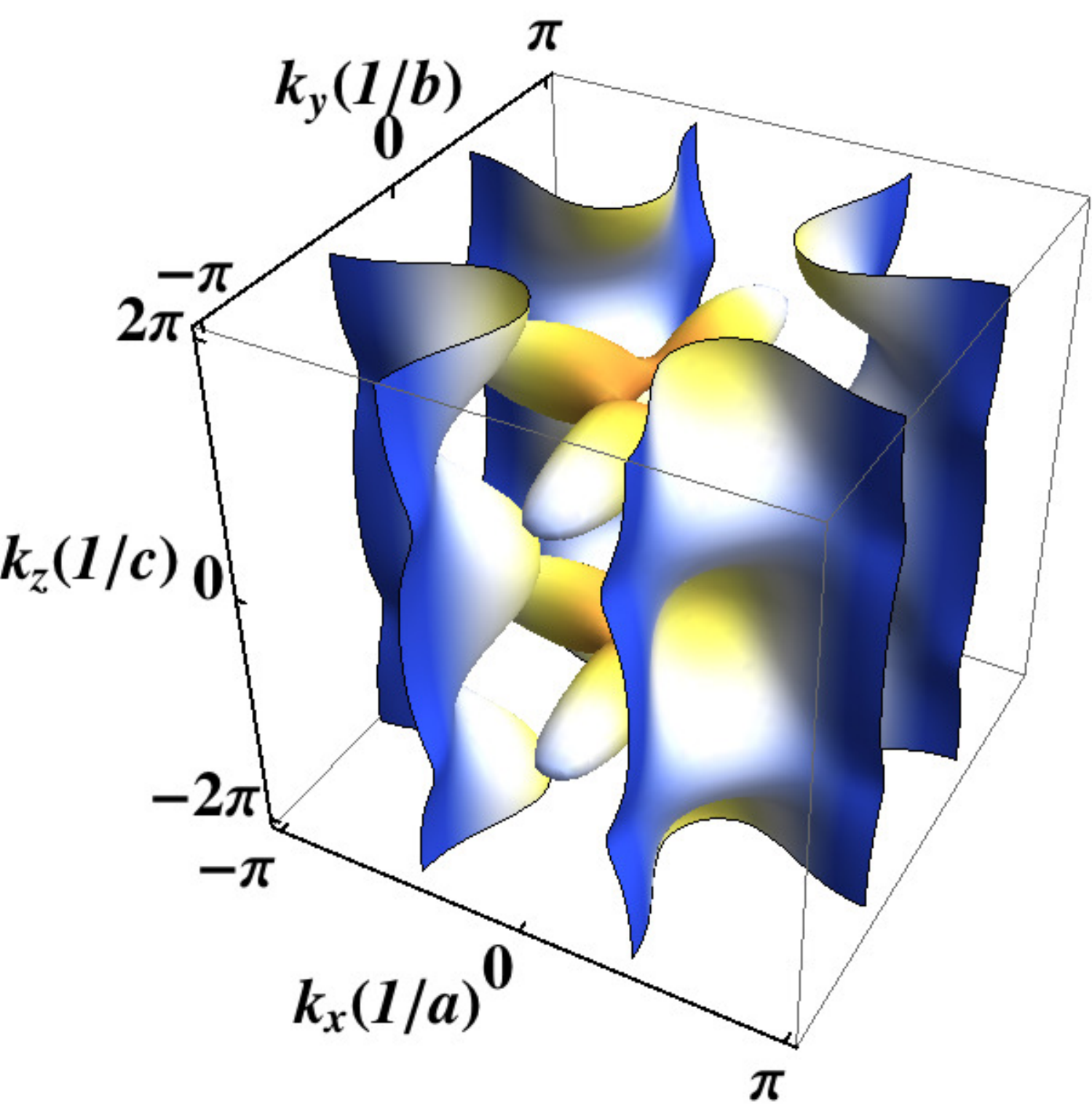}
 %       \label{fig:first_sub}
    }
    \subfigure[]
    {
        \includegraphics[width=0.46\linewidth]{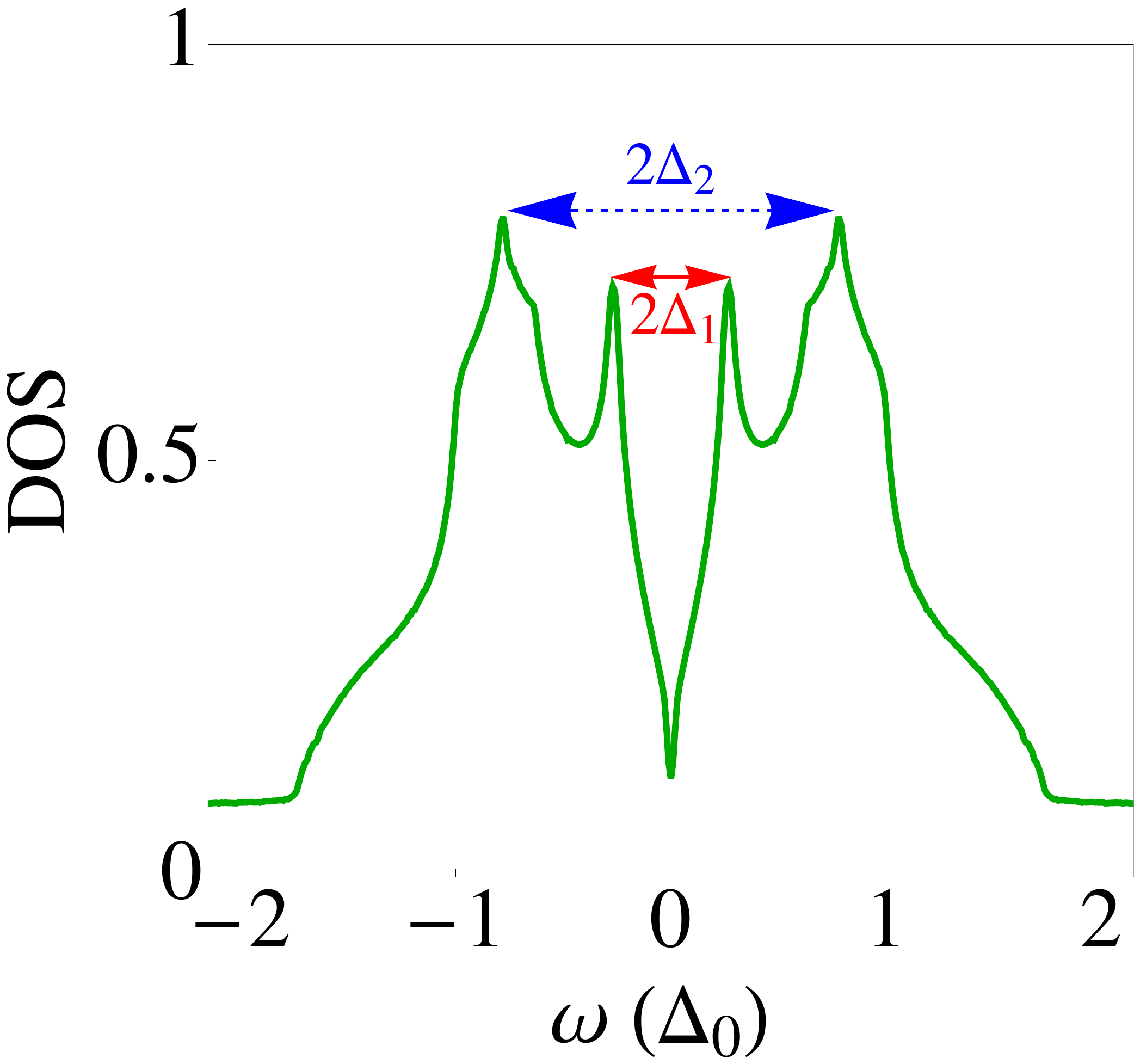}
 %       \label{fig:first_sub}
    }\\
    \subfigure[]
    {
          \includegraphics[width=0.46\linewidth]{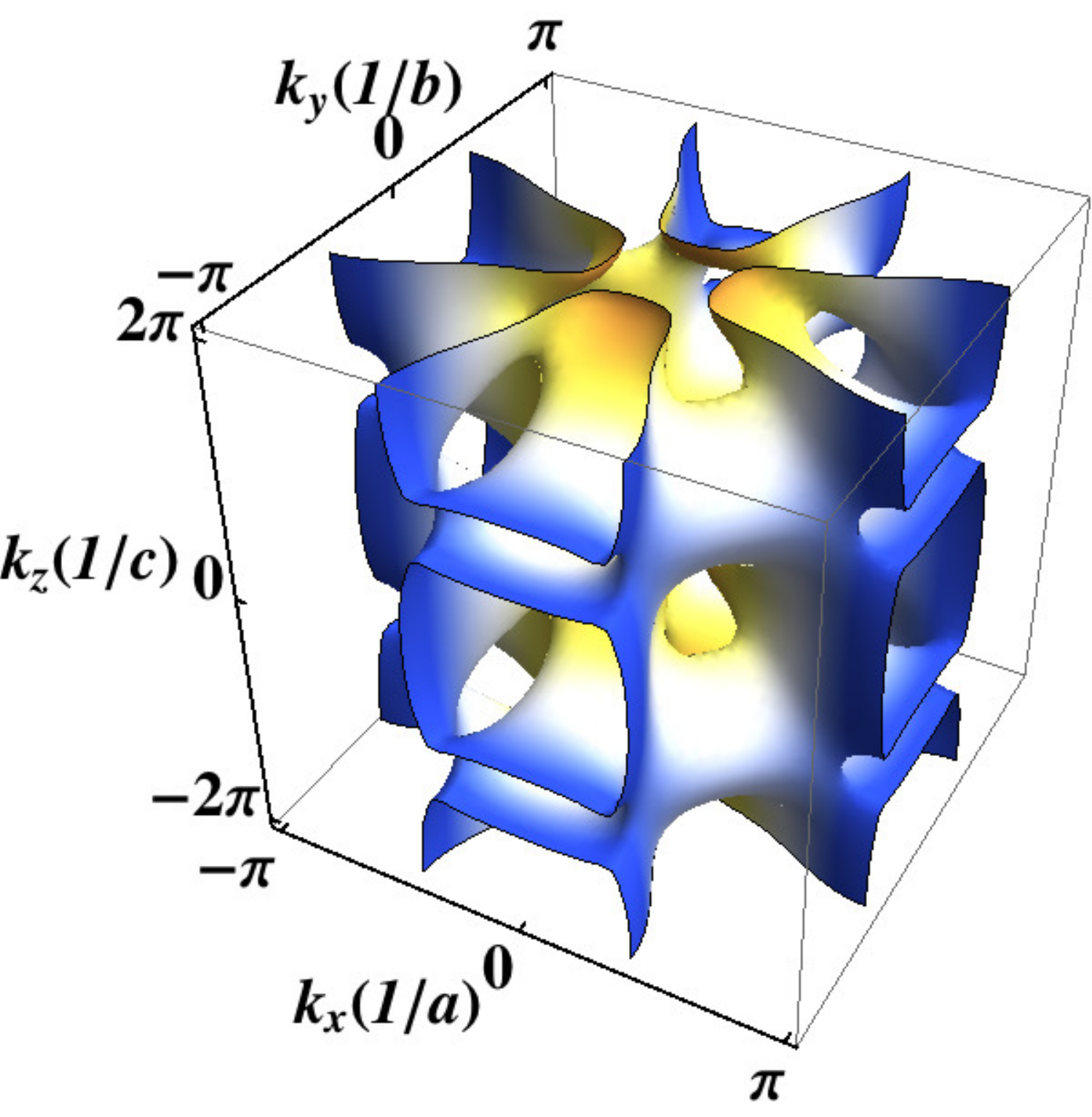}
%        \label{fig:first_sub}
    }
    \subfigure[]
    {     \includegraphics[width=0.46\linewidth]{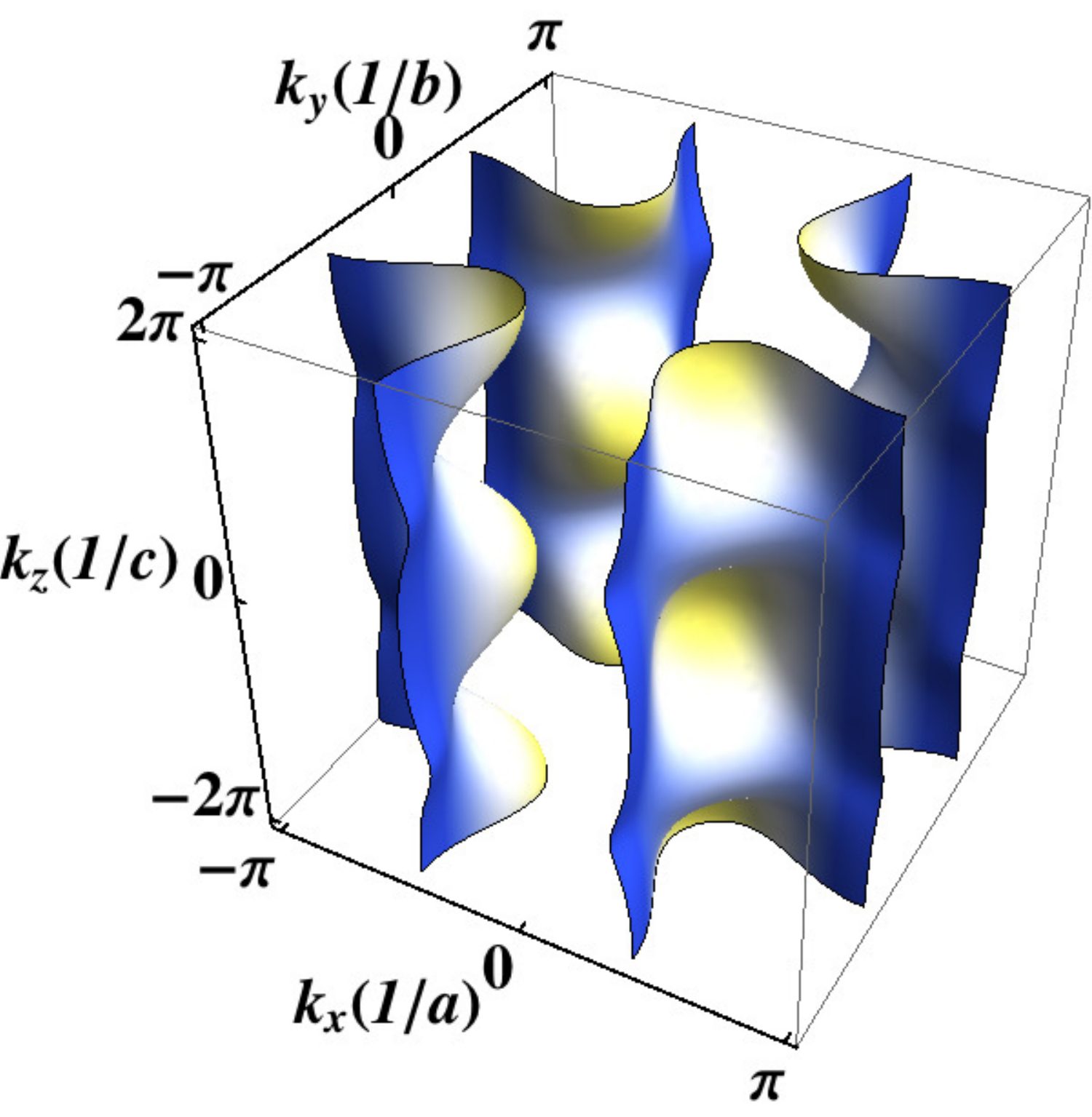}
 %       \label{fig:first_sub}
    }
    \caption{(Color online) Fermi Surface (a) and quasiparticle density of states (b) for  zero magnetic field, here  $2\Delta_1=0.56\Delta_0$ and  $2\Delta_2=1.5\Delta_0$. (c)  and (d) show Fermi surfaces for spin up and spin down,  respectively,  in presence of the  magnetic field $B_0{\hat z}$, corresponding to Zeeman energy splitting $h_{B_0}^f=g^f_{\parallel}\tilde{\chi}\mu_BB_0 =0.3\Delta_0$ which is also used in subsequent figures. 
}
\label{fig1}
\end{figure}
 %%%%%%%%%%%%%%%%%%%%%% figure %%%%%%%%%%%%%%%%%%%%%%%%%%%%%%%%%%%%%%%%%%%%%%%%%%%

The Anderson lattice model  provides a convenient frame to model the heavy quasiparticle bands in $\rm{ CeMIn_5} $ realistically \cite{tanaka:06}. The electronic structure of $\rm{ CeMIn_5}$ has been investigated by using tight binding models with effective hybridisation for f-electrons and c-(conduction) electrons \cite{maehira:03,tanaka:06}. The crystalline electric field (CEF)-splitting is approximately three times larger than the quasiparticle band width ($W\approx 4$ meV), then one may restrict the model  to consider the lowest $\Gamma_7^{(1)}$ Kramers doublet of 4f states \cite{Christianson:04}. This can be described by a pseudo-spin $\sigma = \ua\da$ degree of freedom. Choosing its quantization axis along the ${\hat z}$ direction which is defined by  the magnetic field, the Anderson lattice model Hamiltonian  for the two
hybridized conduction and localized orbitals (c,f)  which are doubly Kramers degenerate is given by
\begin{eqnarray}
{\cal H}
&=&
\sum\limits_{{\bf k}\sigma }
\varepsilon ^c_{{\bf k}\sigma}c_{{\bf k}\sigma
}^{\dagger}c_{{\bf k}\sigma }
+
\varepsilon ^f_{{\bf k}\sigma} f_{{\bf k}\sigma} ^{\dagger}f_{{\bf k}\sigma}
+V_{{\bf k}}\left( c_{{\bf k}\sigma }^{\dagger}f_{{\bf k}\sigma}
+h.c.\right)
 \nonumber \\
&&+\sum\limits_{{\bf k} {\bf k}^\prime } U_{ff}f_{{\bf k}\uparrow }^{\dagger}f_{{\bf k}\uparrow }f_{{\bf k}^\prime\downarrow}^{\dagger}f_{{\bf k}^\prime\downarrow }.
\label{eq-}
\end{eqnarray}
Where $c_{{\bf k}\sigma
}^{\dagger}$ creates an electron with spin $\sigma$
in the conduction orbital  with wave vector ${\bf k}=(k_x,k_y,k_z)$.
Furthermore, 
$\varepsilon^c_{{\bf k}\sigma}=\epsilon^{c}_{{\bf k}}-{\cal H}_B^c$ and $\varepsilon^f_{{\bf k}\sigma}=\epsilon^f_{{\bf k}}-{\cal H}_B^f$ , 
where $\epsilon^{c}_{{\bf k}}$ and $\epsilon^{f}_{{\bf k}}$  are effective tight binding
dispersions of the conduction band and the renormalized dispersion for the $f$ band, respectively. The  Zeeman splittings of bands due to the effective molecular fields are given by ${\cal H}_B^c=h_B^c\sigma_z$ with $h_B^c=\frac{1}{2}g^{c}\tilde{\chi}\mu_B B$ and ${\cal H}_B^f=h_B^f\sigma_z$ with  $h_B^f=\frac{1}{2}g_{\alpha}^{f}\tilde{\chi}\mu_BB$. Here 
$\sigma_z$ is the Pauli matrix, $B$ is the magnetic field, $g^{c} $ is the g-factor for $c$-electrons,
 $g_{\alpha}^{f}$ the g-factor for $f$-electrons in direction $\alpha =\parallel,\perp$ with respect to the tetragonal plane, $\tilde{\chi}$ the Stoner enhancement factor of the homogeneous susceptibility due to quasiparticle interactions and  $\mu_B $ is the Bohr magneton. The anisotropy of the g-factors is obtained from that of magnetization or spin susceptibility \cite{Tayama:2002} as $g^f_\perp/g^f_\parallel =2.3$ assuming that the f-electron contribution in the magnetization dominates. For that reason we chose a small $g^c/g^f_\parallel =0.2$. Furthermore $f_{{\bf k}\sigma} ^{\dagger}$ creates the f-electron with momentum \bk~ and pseudo spin $\sigma$, and $U_{ff}$ is its on-site Coulomb repulsion. Finally $V_{{\bf  k}}$ is the  hybridization energy between the lowest 4f doublet
and conduction bands which contains implicitly the effect of spin orbit and the  CEF term and is taken as momentum independent.

 %%%%%%%%%%%%%%%%%%%%%% figure %%%%%%%%%%%%%%%%%%%%
\begin{figure}
\centering
\includegraphics[width=0.94\linewidth]{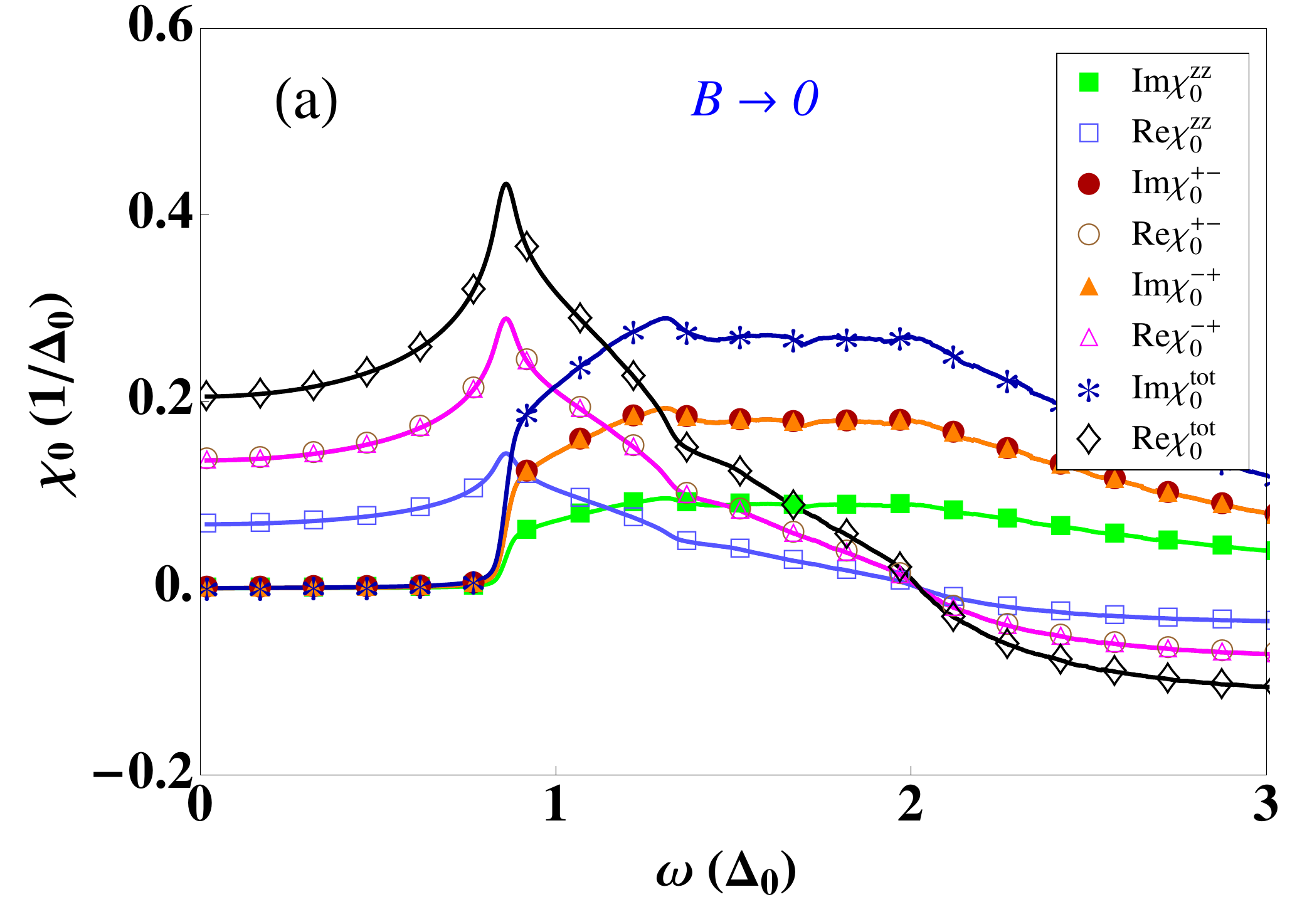}
\\
\includegraphics[width=0.94\linewidth]{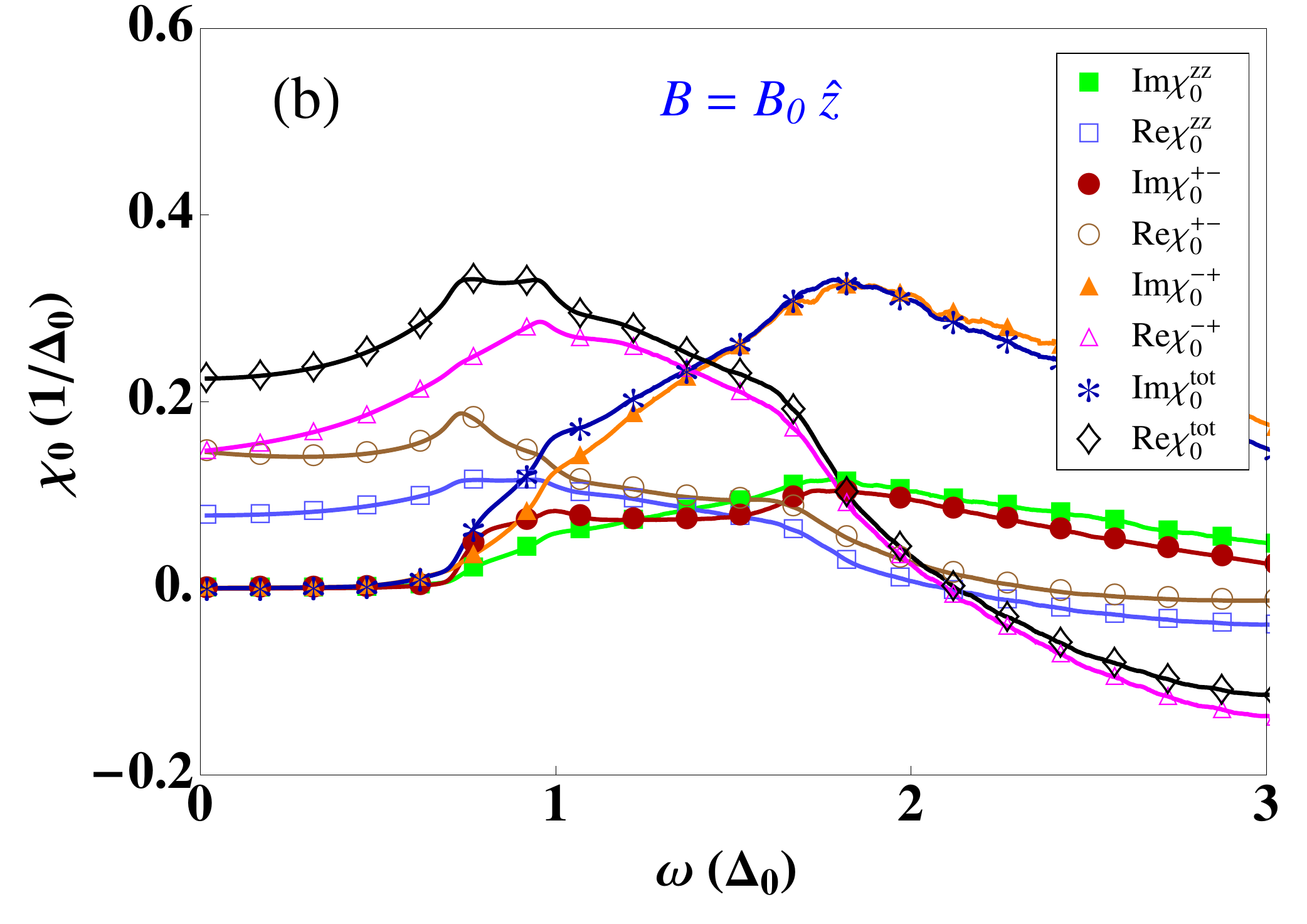}
\\
\includegraphics[width=0.94\linewidth]{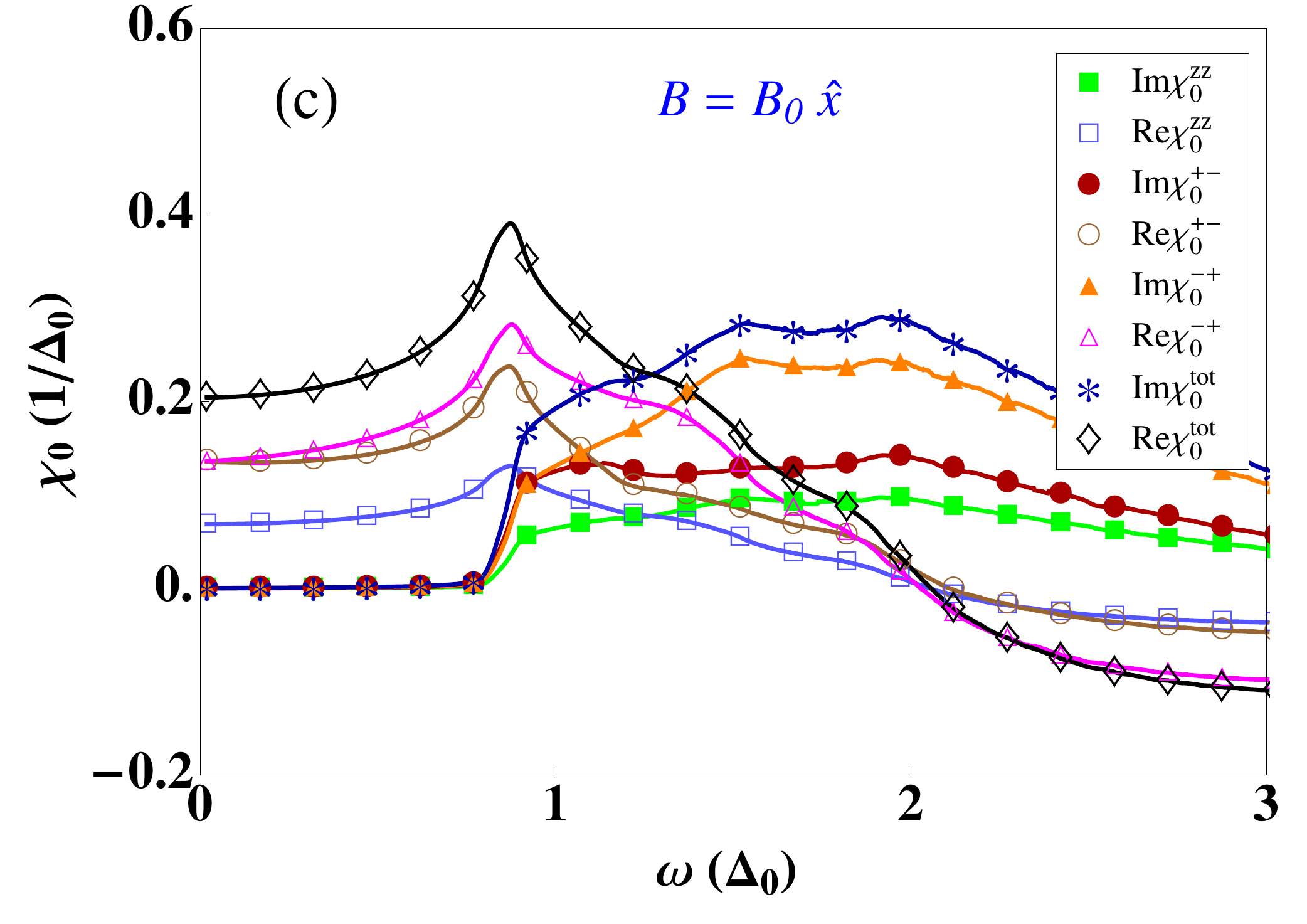}

\caption{(Color online) Individual compounds ($\chi^{ll'}_{0}({\bf Q},\omega)$) and total ($\chi^{tot}_{0}({\bf Q},\omega)$) of the unperturbed pseudo spin susceptibility in absence (a) and at presence  of the magnetic field along the ${\hat z}$  direction (b), and for magnetic field along the ${\hat x}$ direction (c). ($B_0$ is the same  as  Fig.\ref{fig1}).
}
%}
%
\label{fig2}
\end{figure}
%%%%%%%%%%%%%%%%%%%%%%%%%%%%%%%%%%%%%%%%%%%%%
 
It is known \cite{tanaka:06} that in the limit of  $U_{ff}\rightarrow \infty$ where double occupation  of the f-states are excluded,
 an  auxiliary boson defines the mean field (MF)  Hamiltonian as\cite{Akbari:11,Akbari:12}
\begin{eqnarray}
{\cal H}
&=&
\sum\limits_{{\bf k}\sigma }
\varepsilon ^c_{{\bf k}\sigma}c_{{\bf k}\sigma
}^{\dagger}c_{{\bf k}\sigma }
+
\tilde{\varepsilon}^f_{{\bf k}\sigma} f_{{\bf k}\sigma} ^{\dagger}f_{{\bf k}\sigma}
+\tilde{V}_{{\bf k}}\left( c_{{\bf k}\sigma }^{\dagger}f_{{\bf k}\sigma}
+h.c.\right)
 \nonumber \\
 &&+\lambda(r^2-1).
\end{eqnarray}
The MF Hamiltonian can be diagonalized using
the unitary transformation, 
\bea
f_{{\bf k}\sigma }=u_{+, {\bf k} \sigma} \alpha_{+,{\bf k}\sigma }+u_{-, {\bf k} \sigma} \alpha_{-,{\bf k}\sigma }
\nonumber\\
c_{{\bf k}\sigma }=u_{-, {\bf k} \sigma} \alpha_{+,{\bf k}\sigma }-u_{+, {\bf k} \sigma} \alpha_{-,{\bf k}\sigma }
\eea
and as a result one can find the MF quasiparticle Hamiltonian as
\bea
{\cal H}^{\rm MF}&=& \sum\limits_{{\pm},{\bf k}\sigma }
E^{\pm}_{{\bf  k}\sigma}\alpha^{\dagger}_{\pm,{\bf k}\sigma}\alpha_{\pm,{\bf k}\sigma},%
+\lambda(r^2-1),
 \eea
 where the two pairs of  quasiparticle bands (pairwise degenerate for zero field)  are given by
 \bea
E^{\pm} _{{\bf  k}\sigma}&=&\frac{1}{2}
\biggl[
\varepsilon^{c}_{{\bf  k}\sigma}+\tilde{\varepsilon}^{f}_{{\bf  k}\sigma}\pm\sqrt{(\varepsilon^{c}_{{\bf  k}\sigma}-\tilde{\varepsilon}^{f}_{{\bf  k}\sigma})^2+4\tilde{V}^2_{{\bf  k}}}
\biggr],
\eea
and the quasiparticle mixing amplitudes are obtained from
\bea
u_{\pm, {\bf k} \sigma}^2 =\frac{1}{2}
\left( 
1\pm \frac{\varepsilon^{c}_{{\bf  k} \sigma}-\tilde{\varepsilon}^{f}_{{\bf  k} \sigma}}{\sqrt{(\varepsilon^{c}_{{\bf  k} \sigma}-\tilde{\varepsilon}^{f}_{{\bf  k} \sigma})^2+4\tilde{V}^2_{{\bf  k}}}}
\right).
\eea

Using the parameters defined in Ref. \onlinecite{tanaka:06} for the above quasiparticle band structure 
we plot the corresponding  FS in Fig.\ref{fig1}(a) in the  absence of a magnetic field. 
Since the Fermi level is located in the lower 
band, $E^-_{{\bf  k}\sigma}$,
we can neglect the upper band, $E^+_{{\bf  k}\sigma}$ for discussing the low energy spin excitations. 

%%%%%%%%%%%%%%%%%%%%%%%%%%%%%%%% figure %%%%%%%%%%%%%%%%%%%%
\begin{figure}
\centering
\includegraphics[width=0.97\linewidth]{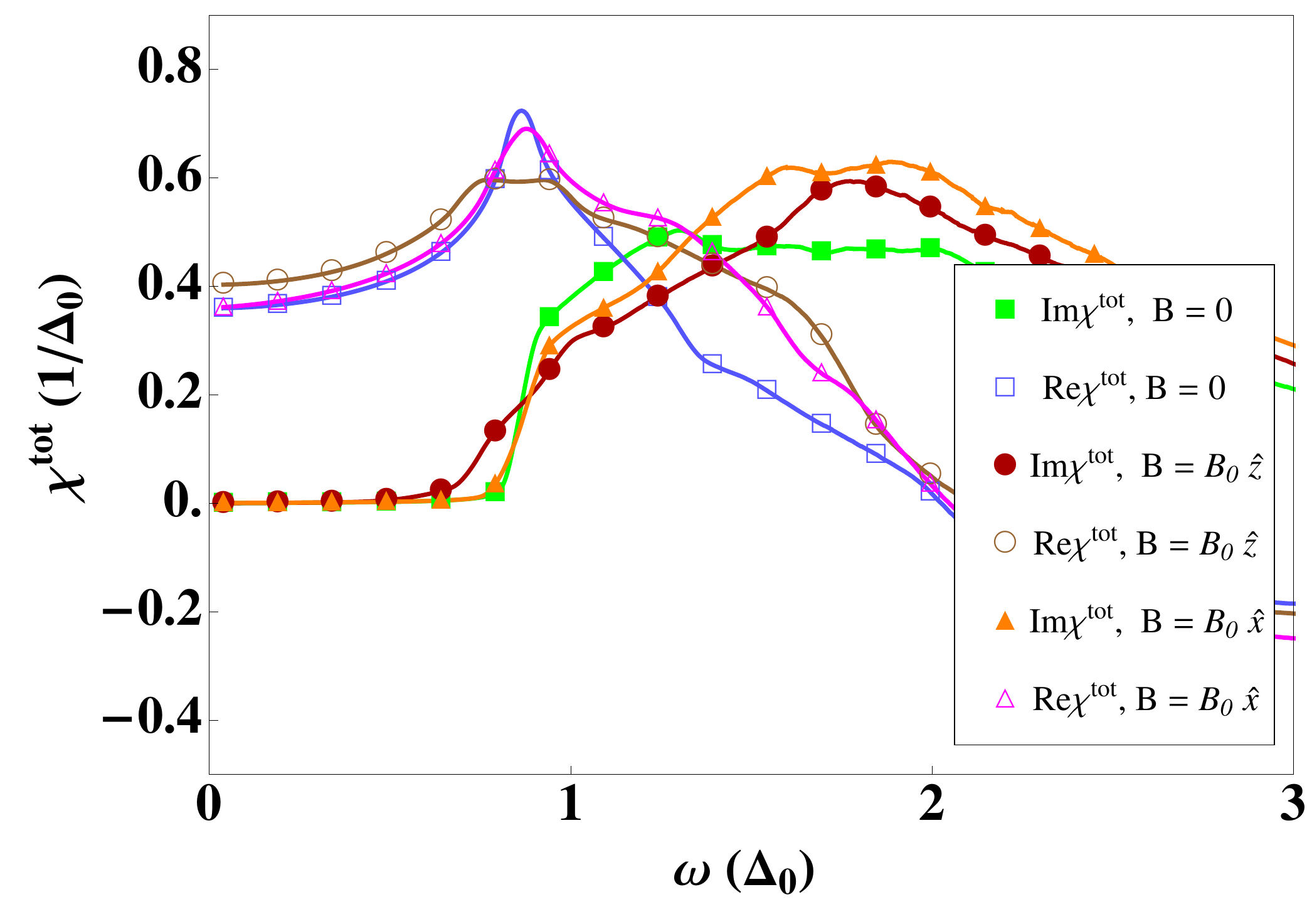}
\caption{(Color online) The total unperturbed physical susceptibility (including matrix elements $m_\parallel, m_\perp$;
$\chi^{tot}({\bf Q},\omega)= \chi^{zz}+ \frac{1}{2}[
  \chi^{+-}+ \chi^{-+}]$)  for zero field and finite field along ${\hat z}$, and ${\hat x}$  directions. 
($B_0$ is the same  as  Fig.\ref{fig1}).
}
%}
%
\label{fig3}
\end{figure}
%%%%%%%%%%%%%%%%%%%%%%%%%%%%%%%%%%%%%%%%%%%%%%

The superconducting pairs are then formed from the heavy quasiparticles of the lower band leading to a pairing potential in the  Hamiltonian according to
\be
{\cal H^{\rm SC}}=\sum\limits_{{\bf k}}\Delta_{{\bf k}}\left(
\alpha_{-,{\bf k}\uparrow }^{\dagger} \alpha_{-,-{\bf k}\downarrow }^{\dagger}+h.c.\right),
\ee
where $\Delta_{{\bf k}}$ is the superconducting d-wave gap function for CeMIn$_5$ given by
\be
\Delta_{{\bf k}}=\frac{\Delta_{0}}{2}(\cos k_x- \cos k_y).
\label{d-wave}
\ee
By defining  the new Nambu spinors as
$\hat{\psi}^{\dagger}_{\bf k}=
(\hat{\phi}^{\dagger}_{1{\bf k}},\hat{\phi}^{\dagger}_{2{\bf k}})
$, 
the effective Hamiltonian can be written as
\begin{eqnarray}
{\cal H}^{\rm eff} =
\frac{1}{2}
\sum\limits_{{\bf k} }
\hat{\psi}_{{\bf k} }^{\dagger}\hat{\beta}_{{\bf k}}
\hat{\psi}_{{\bf k} },
\label{eq:heff}
\end{eqnarray}
here $\hat{\phi}^{\dagger}_{1{\bf k}}=(
\alpha^{\dagger}_{-,{\bf k} \uparrow},
\alpha ^{\dagger}_{-,{\bf k} \downarrow}
)
$,
and
$\hat{\phi}^{\dagger}_{2{\bf k}}=(
\alpha_{-,-{\bf k} \uparrow},
\alpha_{-,-{\bf k} \downarrow}
)
$
and
\bea
\hat{\beta}_{{\bf k}}=
\left[
 \begin{array}{cccc}
E^-_{{\bf  k} \uparrow}  & 0 & 0 &\Delta_{{\bf k}}\\
0 & E^-_{{\bf  k} \downarrow}&- \Delta_{{\bf k}}  & 0 \\
 0 &-\Delta_{{\bf k}}&- E^-_{-{\bf  k} \uparrow}  & 0 \\
\Delta_{{\bf k}} & 0 & 0 & - E^-_{-{\bf  k} \downarrow} 
\end{array}
\right].
\eea

%%%%%%%%%%%%%%%%%%%%%% figure %%%%%%%%%%%%%%%%%%%%
\begin{figure}
\centering
\includegraphics[width=0.94\linewidth]{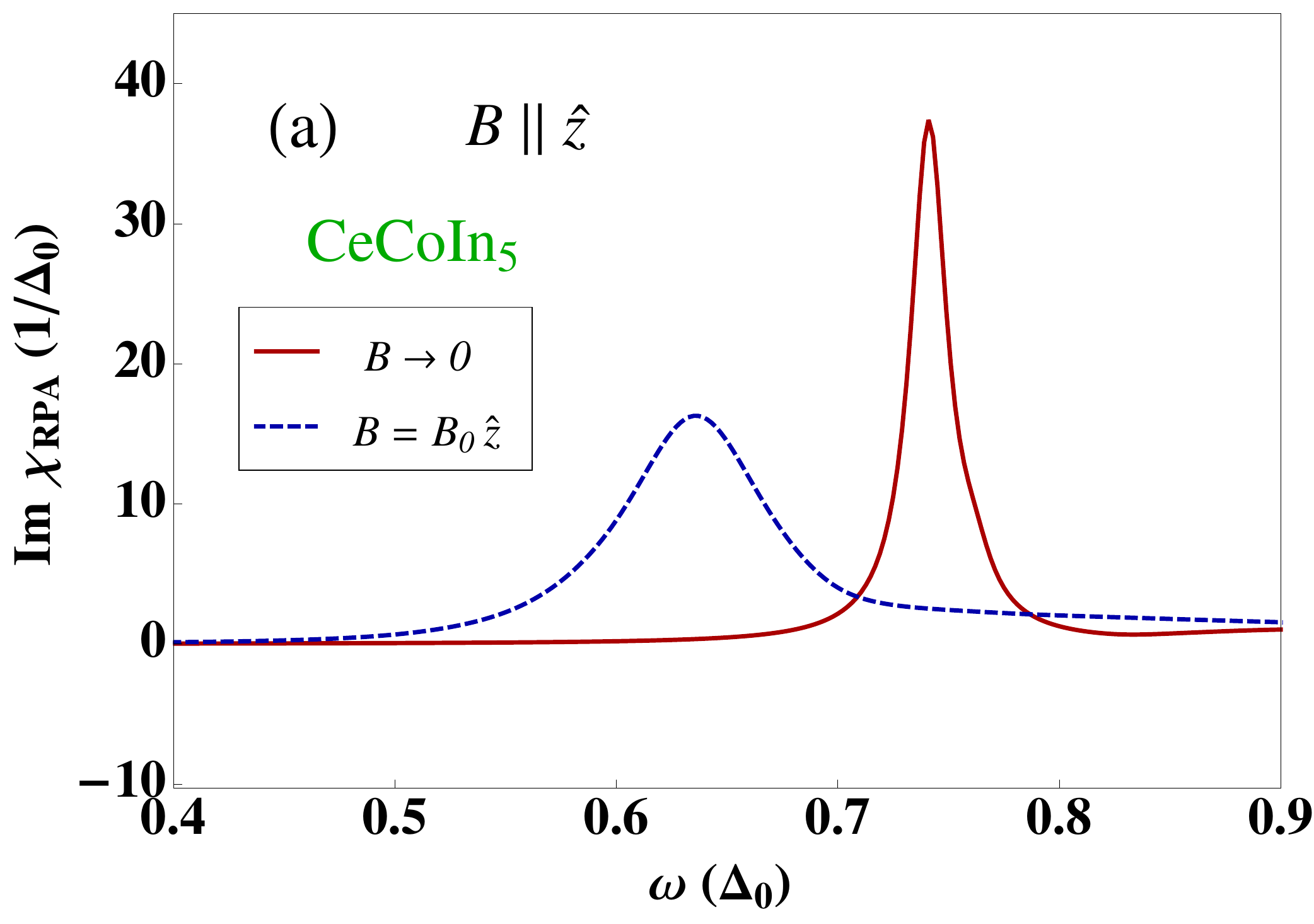}\\
\includegraphics[width=0.94\linewidth]{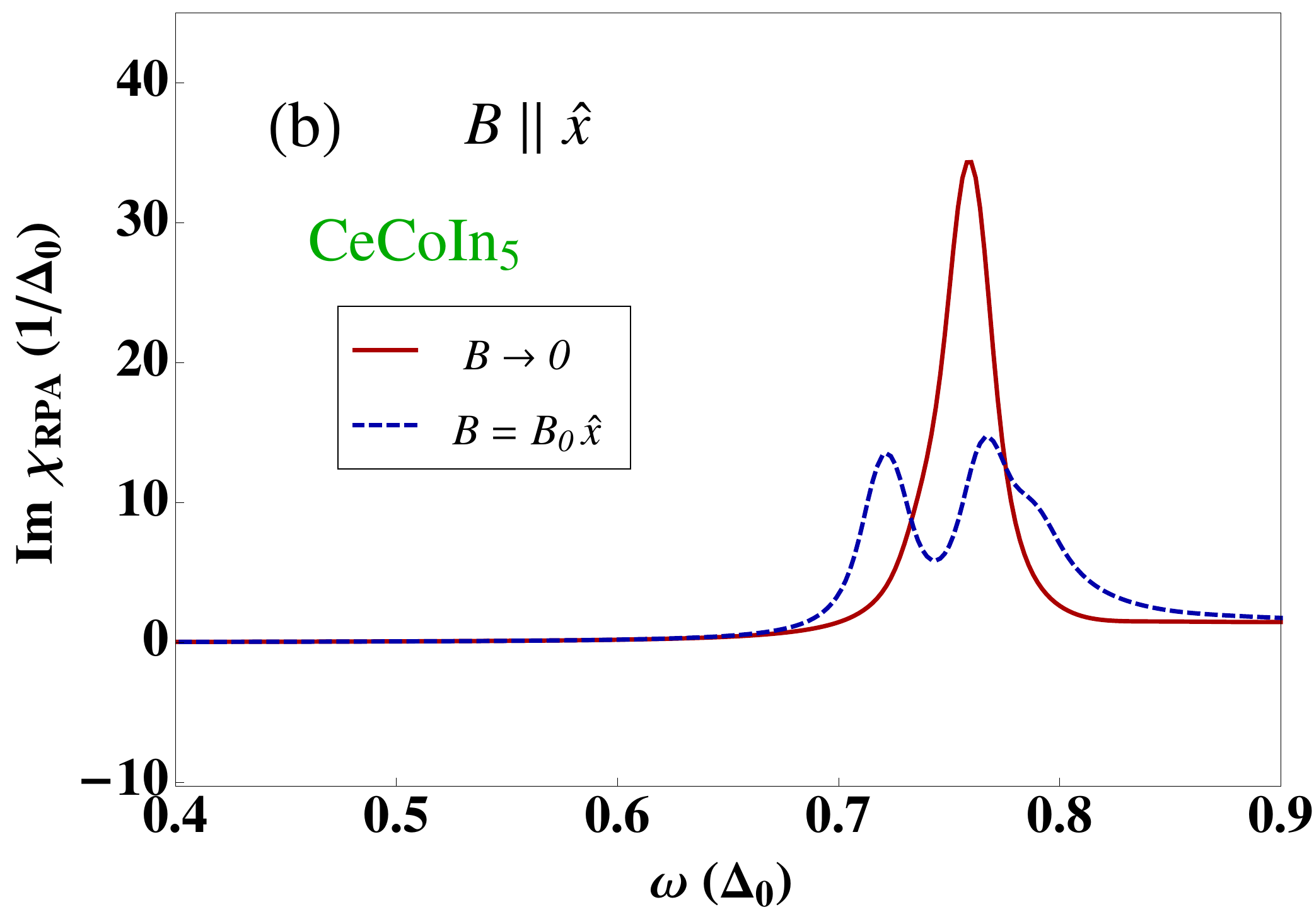}
\caption{(Color online) RPA susceptibility in absence and presence of the magnetic field: along the ${\hat z}$ direction (a), and along the ${\hat x}$ direction (b).
Parameters corresponding to the behaviour as observed in  \Ce with $g_f^\perp/g_f^\parallel=2.3$ and  $J^{\bot}_{{\bf Q}}=2.6\Delta_0$; $J^{\parallel}_{{\bf Q}}= 13.5\Delta_0$.
($B_0$ is the same  as  Fig.\ref{fig1}).
}
\label{fig4}
\end{figure}
%%%%%%%%%%%%%%%%%%%%%%%%%%%%%%%%%%%%%%%%%%%%%
 
%%%%%%%%%%%%%%%%%%%%% figure %%%%%%%%%%%%%%%%%%%%
\begin{figure}
\centering
\includegraphics[width=0.9\linewidth]{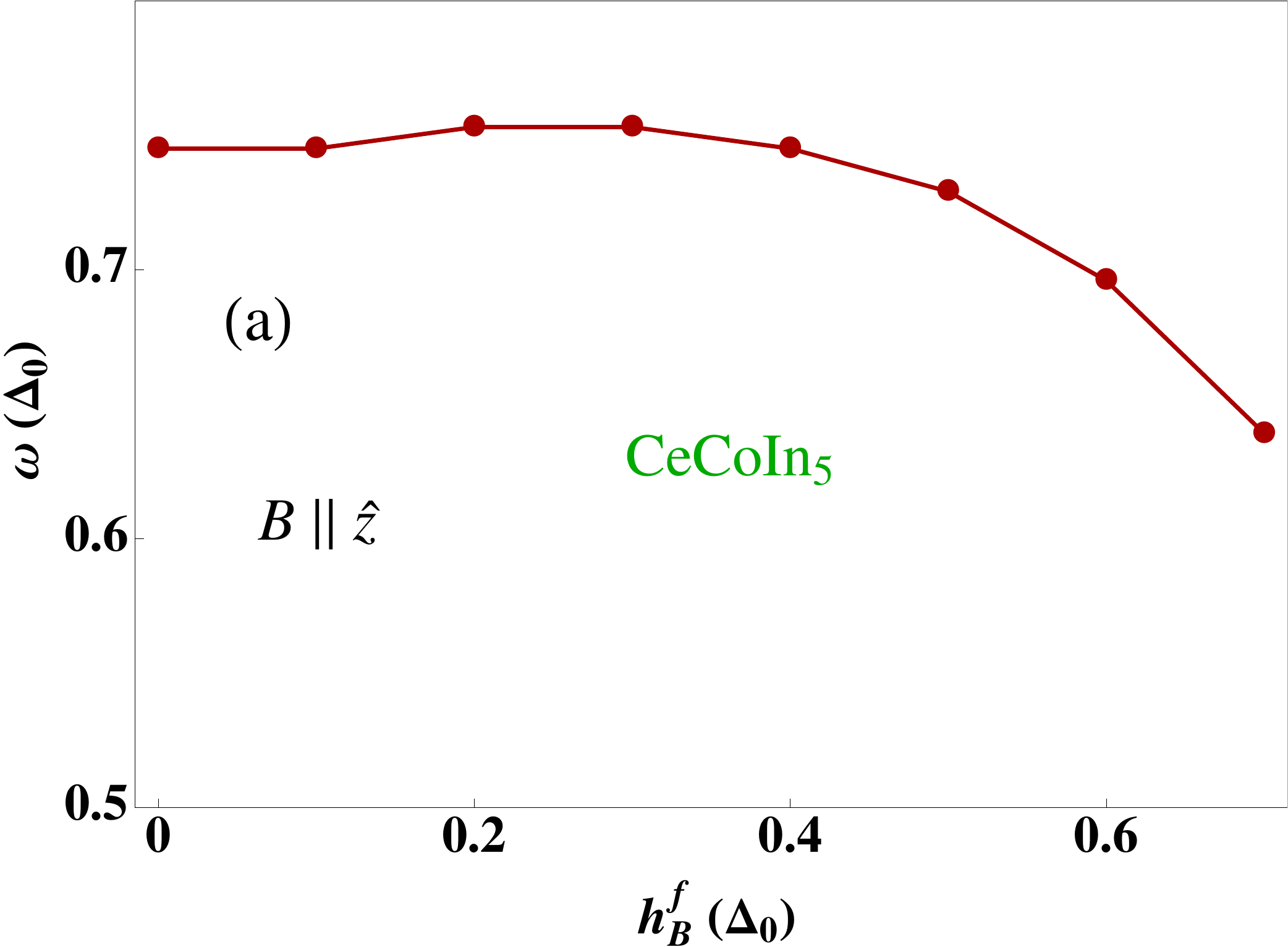}\\
\includegraphics[width=0.9\linewidth]{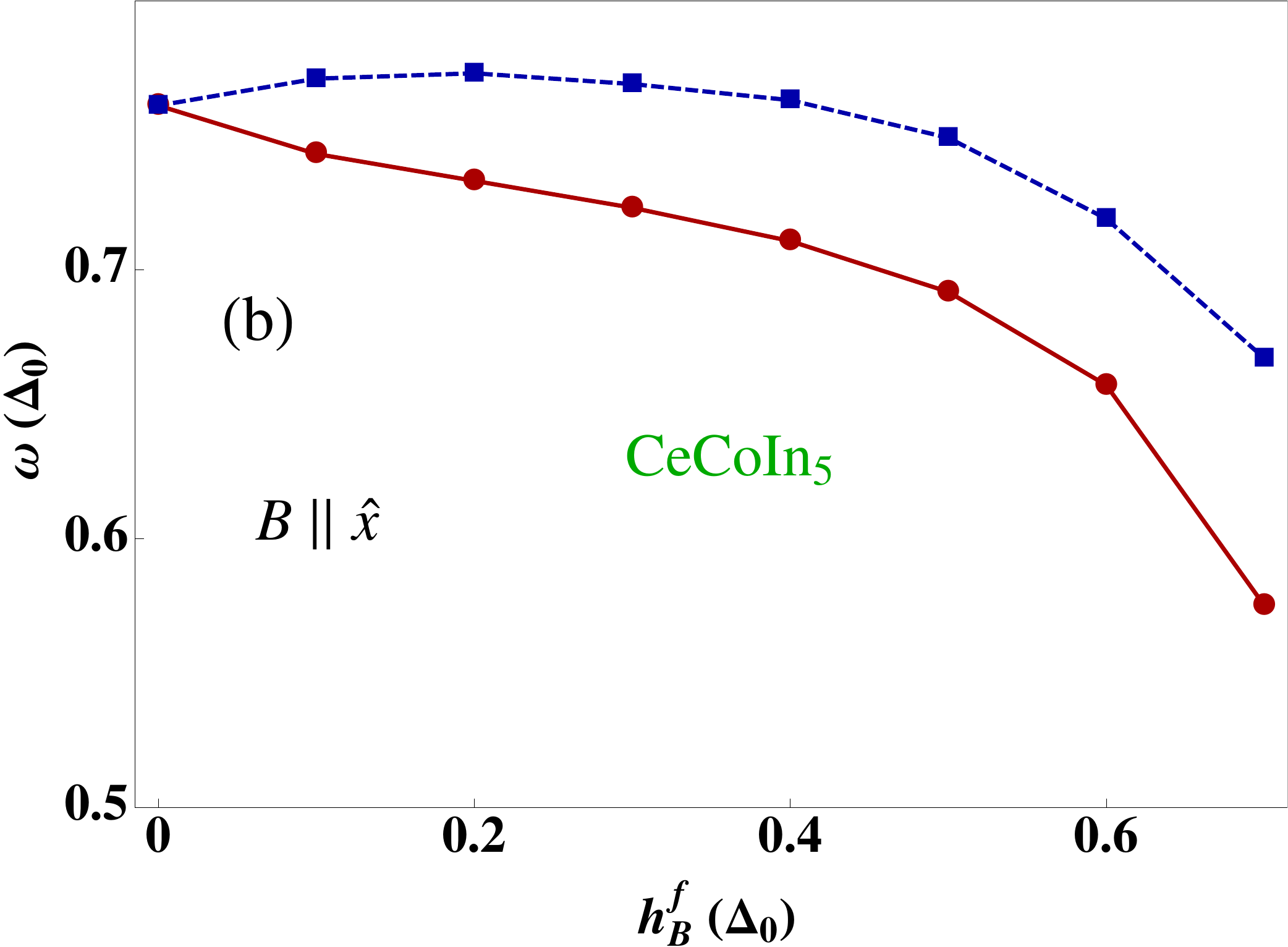}
\caption{(Color online) RPA susceptibility peak positions corresponding to spin resonance energies as function of field strength: along the ${\hat z}$ direction (a), and along the ${\hat x}$ direction (b). Anisotropic g- factors with  $g_f^\perp/g_f^\parallel=2.3$ and quasiparticle interaction parameters  $J^{\bot}_{{\bf Q}}=2.6\Delta_0$; $J^{\parallel}_{{\bf Q}}= 13.5\Delta_0$ leading to similar behaviour as in \Ce~ for both field directions. Here the maximum $h_B^f=0.7\Delta_0$ corresponds to a field $B=39.7/(g_f^l\tilde{\chi})$ in Tesla for direction $l=\perp, \parallel$.}
%}
%
\label{fig5}
\end{figure}
%%%%%%%%%%%%%%%%%%%%%%%%%%%%%%%%%%%%%%%%%%%%%

The propagator matrix of the conduction electrons 
in terms of the Nambu spinor in the Matsubara representation is obtained as
$\hat{G}( {\bf  k},\tau)=-\langle T \hat{\psi}_{{\bf k} }(\tau)
\hat{\psi}_{{\bf k}}^{\dagger}(0)\rangle$.  Using a standard equation of motion  method
one can find that
 \be
 \hat{G}({\bf k},\omega_n)=\left( i\omega_n-\hat{\beta}_{{\bf k}} \right)^{-1}.
 \ee
Explicitly written,
\bea
\hat{G}( {\bf  k},\tau)=
\left[
 \begin{array}{cc}
 \hat{G}_{0}^{11}( {\bf  k},\tau) & \hat{G}_{0}^{12}( {\bf  k},\tau) \\
 \hat{G}_{0}^{21}( {\bf  k},\tau) &  \hat{G}_{0}^{22}( {\bf  k},\tau) 
\end{array}
\right]
\eea
where 
$\hat{G}^{\varsigma \varsigma ^\prime}_{0}( {\bf  k},\tau)=-\langle T \hat{\phi}_{\varsigma{\bf k} }(\tau)
\hat{\phi}_{\varsigma ^\prime{\bf k}}^{\dagger}(0)\rangle$ and $\varsigma, \varsigma'=1,2$  denote the spinor 
components for $\pm{\bf k}$ below Eq.~(\ref{eq:heff}). 

 %%%%%%%%%%%%%%%%%%%%%%%%%%%%%%%%%%%%%%%%%%%%%%%%%
 %%%%%%%%%%%%%%%%%%%%%%%%%%%%%%%%%%%%%%%%%%%%%%%%%
 
 \subsection{Magnetic susceptibility}
 The noninteracting or bare dynamical f-electron  susceptibility  is defined by
 \bea
 \chi_{{\bf q}}^{ll^\prime}(\tau)=-
 \theta(\tau)
 \bra
 T j_{{\bf q}}^{l}(\tau)j_{-{\bf q}}^{l^\prime}(0)
 \ket,
 \eea
 where
 \bea
  j_{{\bf q}}^{l}=\sum\limits_{{\bf k} \sigma \sigma ^\prime} 
  f_{{\bf k}+{\bf q} \sigma} ^{\dagger}
  {\hat M}^{l}_{\sigma \sigma ^\prime} 
  f_{{\bf k} \sigma ^\prime}.
 \eea
 The hybridizing 4f states are in a $|\Gamma_7^{(1)}\rangle$ Kramers doublet state and therefore their physical moment operator 
 (in units of $\mu_B$) can be presented by a  pseudo spin matrix ${\boldsymbol\sigma}$ according to $(l=+,-,z)$
 \be
 \label{moments}
  {\hat M}_l=m_l \sigma_l
  \ee
 Here  $m_l$ are matrix elements and their anisotropy $(m_\pm=m_\parallel; m_z=m_\perp)$  may be directly obtained from that of the experimental spin susceptibility of quasiparticles \cite{Tayama:2002} according to $m_\perp/m_\parallel= g^f_\perp/g^f_\parallel=(\chi^s_\perp/\chi^s_\parallel)^\frac{1}{2} = 2.3$. Their absolute value is not significant as it only sets the overall scale. Then,  the bare physical moment susceptibility can be expressed as
 \be
 \chi_{}^{ll^\prime}({\bf q},\omega)=m_l  m_{l '} \chi_{0}^{ll^\prime}({\bf q},\omega),
 \label{Sus0}
 \ee
 where $ \chi_{0}^{ll^\prime}({\bf q},\omega)$ is the pseudo spin susceptibility defined by
  \bea\nonumber %&&
 \chi_{0}^{ll^\prime}({\bf q},\omega)=
 \sum\limits_{{\bf k},\{ \sigma \}} 
   {\boldsymbol\sigma}^{l}_{\sigma \sigma ^\prime}   {\boldsymbol\sigma}^{l^\prime}_{\sigma_1^{}\sigma_1^\prime  } 
   u_{-,{\bf k}+{\bf q}\sigma_{}^{} }   u_{-,{\bf k}+{\bf q}\sigma_{1}^{\prime} } 
   u_{-,{\bf k} \sigma_{1}^{} }
   u_{-,{\bf k} \sigma_{}^{\prime} }
   \\\nonumber %&&
    \int d\omega^\prime
   {\hat G}^{11}_{0\sigma\sigma_1^\prime}({\bf k}+{\bf q},\nu+\omega^\prime)      {\hat G}^{11}_{0\sigma_1^{} \sigma ^\prime}({\bf k},\omega^\prime)
  \mid _{i\nu\rightarrow \omega+i 0^+}.
     \\%&&
  \eea
 \\
With the definition of  the new basis as ${\hat \varphi}^{\dagger}_{{\bf k}}=(  b_{+,1{\bf k}} , b_{+,2{\bf k}} , b_{-,2{\bf k}} , b_{-,1{\bf k}} )$ and applying the Bogoliubov  transformation given by
 \bea
\alpha ^{\dagger}_{-,{\bf k} \uparrow}=v_{+,1{\bf k}} b_{+,1{\bf k}} +v_{-,1{\bf k}} b_{-,1{\bf k}} 
\nonumber\\
\alpha_{-,-{\bf k} \downarrow}=-v_{-,1{\bf k}} b_{+,1{\bf k}} +v_{+,1{\bf k}} b_{-,1{\bf k}} 
\nonumber\\
\alpha ^{\dagger}_{-,{\bf k} \downarrow}=v_{+,2{\bf k}} b_{+,2{\bf k}} -v_{-,2{\bf k}} b_{-,2{\bf k}} 
\nonumber\\
\alpha_{-,-{\bf k} \uparrow}=v_{-,2{\bf k}} b_{+,2{\bf k}} +v_{+,2{\bf k}} b_{-,2{\bf k}} 
\eea
the effective Hamiltonian in Eq.~(\ref{eq:heff}) is diagonalized by
$
 \hat{\beta}_{{\bf k}}^{d}=P_{{\bf k}}  \hat{\beta}_{{\bf k}} P^{-1}_{{\bf k}}
$,
 where $P_{{\bf k}}$ is a $4\times 4$ matrix, composed of the eigenvectors of $ \hat{\beta}_{{\bf k}}$. Here $ \hat{\beta}_{{\bf k}}^{d}$ is the diagonal matrix constructed from the corresponding eigenvalues, 
  \be
{\bar E}^{\pm} _{1{\bf  k}}={\bar E}^{\pm} _{2,-{\bf  k}}=\frac{1}{2}
\biggl[
E^-_{{\bf  k} \uparrow} -E^-_{-{\bf  k} \downarrow} \pm\sqrt{(E^-_{{\bf  k} \uparrow} +E^-_{-{\bf  k} \downarrow})^2+4\Delta^2_{{\bf  k}}}
\biggr],
\ee
and $P^{-1}_{{\bf k}}$ is the matrix inverse of $P_{{\bf k}}$. Furthermore,
\be
v_{\pm,1 {\bf k}}^2 =v_{\pm,2, -{\bf k}}^2 =\frac{1}{2}
\left( 
1\pm \frac{E^-_{{\bf  k} \uparrow} +E^-_{-{\bf  k} \downarrow} }{\sqrt{(E^-_{{\bf  k} \uparrow} +E^-_{-{\bf  k} \downarrow} )^2+4\Delta^2_{{\bf  k}}}}
\right).
\ee
\\
Then
$
 {\hat G}^{d}_{}({\bf k},\omega_n)=\left( i\omega_n-\hat{\beta}^{d}_{{\bf k}} \right)^{-1}
$
 is diagonal and we can write
 \bea
 {\hat G}^{11}_{0\sigma\sigma_1}({\bf k},\omega_n) 
   &=&
  \sum\limits_{s^\prime=1} ^{4}  \gamma_{\sigma \sigma_1 s^\prime }^{{\bf k}} {\hat G}^{d} _{s^\prime}({\bf k},\omega_n),
 \eea
 where
$\gamma_{\sigma \sigma_1s^\prime }^{{\bf k}}=P^{-1}_{{\bf k} \sigma s^\prime} P_{{\bf k}s^\prime \sigma_1}.
$
Thus, the pseudo-spin susceptibility can now be expressed as
  \bea
  \label{Chi0}
&& \chi_{0}^{ll^\prime}({\bf q},\omega)=
 \sum\limits_{{\bf k},\{ \sigma \}} 
 \sum\limits_{\{s \}} 
   {\boldsymbol\sigma}^{l}_{\sigma \sigma ^\prime}   {\boldsymbol\sigma}^{l^\prime}_{\sigma_1^{}\sigma_1^\prime  } 
    \gamma_{\sigma\sigma_1^{\prime}s_2}^{{\bf k}+{\bf q}}  \gamma_{\sigma_1^{}\sigma^\prime_{}s_2^{\prime}}^{{\bf k}}
   \nonumber\\&&%\hspace{1.8cm}
   u_{-,{\bf k}+{\bf q}\sigma_{}^{} }   
   u_{-,{\bf k}+{\bf q}\sigma_{1}^{\prime} } 
   u_{-,{\bf k} \sigma_{1}^{} }
   u_{-,{\bf k} \sigma_{}^{\prime} }  
  % %\hspace{1.81cm}
   %
\frac{f(\hat{\beta}^{d}_{{\bf k}+{\bf q},s_2})-f(\hat{\beta}^{d}_{{\bf k},s_2^\prime} )}
{\omega-(\hat{\beta}^{d}_{{\bf k}+{\bf q},s_2} -\hat{\beta}^{d}_{{\bf k},s_2^\prime} )},
   \nonumber\\&&
\label{susfin}
  \eea
where  $f(\epsilon)$ is the Fermi function. The combination of prefactors in the sum of Eq.~(\ref{susfin}) are the combined coherence factors arising from the hybridisation and the superconducting state.
Finally the cartesian dynamic magnetic susceptibility tensor in random phase approximation (RPA) has the form 
\be
\hat{\chi}_{RPA}({\bf q},\omega)=[1- {\hat J}({\bf q}) \hat{\chi}({\bf q},\omega)]^{-1}\hat{\chi}({\bf q},\omega),
\label{RPAtensor}
\ee
where  ${\hat J}({\bf q})$ is the effective quasiparticle interaction matrix with non-zero elements:
$J^{zz}_{{\bf q}}=J^{\bot}_{{\bf q}}$, $J^{+-}_{{\bf q}}=J^{-+}_{{\bf q}}=J^{\parallel}_{{\bf q}}$. Similar to the g-factors they may
be anisotropic. Furthermore their momentum dependence can be modeled by a Lorentzian function which is peaked at the 
wave vector \bQ~ associated with the indirect hybridization gap \cite{akbari:09}.
Finally the total dynamic magnetic susceptibility is obtained as the trace of the tensor according to
\be
 \chi_{RPA}({\bf q},\omega)= \chi^{zz}_{RPA}({\bf q},\omega)+
 \frac{1}{2}
 \left[
  \chi^{+-}_{RPA}({\bf q},\omega)+ \chi^{-+}_{RPA}({\bf q},\omega)
  \right].
  \label{chiRPA}
\ee
For momentum transfer {\bf q} along ${\bf Q}=\bigl(\frac{1}{2},\frac{1}{2},\frac{1}{2}\bigr)$ direction this is directly proportional to the dynamical structure factor
$S({\bf q},\omega)$ observed in INS experiments.
 
%%%%%%%%%%%%%%%%%%%%%%%%%%%%%%%%%%%%%%%%%%%%%%%%%%%%%%%%%%%%%%%%%%%%%%%%%%
%%%%%%%%%%%%%%%%%%%%%%%%%%     Section III     %%%%%%%%%%%%%%%%%%%%%%%%%%%%
%%%%%%%%%%%%%%%%%%%%%%%%%%%%%%%%%%%%%%%%%%%%%%%%%%%%%%%%%%%%%%%%%%%%%%%%%%

 \section{Numerical Results} 
 \label{secIII}
 
 In this section, we evaluate numerically  the dynamic magnetic susceptibility for our model of 115 systems.
 We consider two different cases, namely magnetic field along ${\hat z}$- direction (out-of-plane: c-axis) and along ${\hat x}$- direction (in-plane: a-axis). For clarity we give first the relevant physical parameters of \Ce~ to simplify comparison with experimental results. For the main (lower) tunneling gap (Fig.~\ref{fig1}) we have $2\Delta_1=0.56\Delta_0=0.92 \mbox{meV}$ \cite{Rourke:2005} or $\Delta_0=1.64 \mbox{meV}$ for the gap amplitude in Eq.~(\ref{d-wave}). The  spin resonance energy for zero field is $\omega_r=0.6$ meV \cite{Stock:08} or $\omega_r/2\Delta_1\simeq 0.65$.
  
 First,  using Eq.(\ref{Chi0}), we calculate the
 individual components $\chi^{ll'}_{0}({\bf Q},\omega)$ of the pseudo-spin susceptibility tensor and its trace $\chi^{tot}_{0}({\bf Q},\omega)= \chi^{zz}_0({\bf Q},\omega)+ \frac{1}{2}[\chi^{+-}_0({\bf Q},\omega)+ \chi^{-+}_0({\bf Q},\omega)]$.
 For $B\rightarrow 0$ Fig.\ref{fig2}.a shows that $\frac{1}{2}\chi_0^{+-}, \frac{1}{2}\chi^{+-}_0$ and $\chi_0^{zz}$ become identical, independent of the field direction due to isotropy in pseudo-spin space. Therefore $\chi^{tot}_{0}$  for $B\rightarrow 0$ in Fig.\ref{fig2}.a is simply three times each of these components. This means that the anisotropy enters only through the matrix elements in the physical moment susceptibility of Eq.~(\ref{Sus0}) discussed below.
The pseudo-spin susceptibility in
presence  of a magnetic field along the ${\hat z}$-direction is shown in Fig.\ref{fig2}.b and for ${\hat x}$-direction in Fig.\ref{fig2}.c.  We note here that for calculating the pseudo spin susceptibility with magnetic field along the  ${\hat x}$-direction, we have applied a $\pi/2$ rotation along the  ${\hat y}$-direction (${\hat x}\rightarrow -{\hat z}$ and ${\hat z}\rightarrow {\hat x}$) and similar in {\bf k}-space.
For finite field and both directions  the individual components for each field direction start to differ because of the polarization of quasiparticle bands in the field. This leads to a change of the nesting conditions of the FS in presence of a magnetic field, shown in Fig.(\ref{fig1}.b), which is different for each susceptibility component.
Most importantly the energy dependence of $ \chi_0^{+-}$ and  $\chi_0^{-+}$ shows a splitting in opposite manner. 
  
We also calculate the total bare physical susceptibility, $\chi^{tot}({\bf Q},\omega)$  resulting from Eq.(\ref{Sus0}) in Fig.\ref{fig3}. 
The anisotropy enters only through the matrix elements $m_\parallel, m_\perp$ in the physical moment susceptibility of Eq.~(\ref{Sus0}). It is shown at zero magnetic field ($B\rightarrow 0$), and also  for magnetic field along ${\hat z}$- and ${\hat x}$-directions (in original axis notation), respectively.
We note that for  $B\rightarrow 0$ the $\frac{1}{2}\chi^{+-}$ and  $\frac{1}{2}\chi^{-+}$ components are still equivalent as in  Fig.\ref{fig2}.a but differ from $\chi^{zz}$ due to the different magnetic moment for in plane and out-of-plane direction (see Eq.\ref{moments}).\\

Finally, from Eq.(\ref{chiRPA}) the interacting RPA  susceptibility and  the spectrum of excitations given by its imaginary part can be calculated. In general form, we obtain
\bea
&& \chi_{RPA}({\bf q},\omega)= 
 \frac{m_z^2\chi^{zz}_{0}}{1-\lambda_z\chi^{zz}_{0}}+
 \nonumber\\ &&
  \frac{(m_x^2+m_y^2) (\chi^{+-}_{0}+\chi^{-+}_{0})-(\lambda_xm_y^2+\lambda_ym_x^2)\chi^{+-}_{0}\chi^{-+}_{0} }
  {4-(\lambda_x+\lambda_y)(\chi^{+-}_{0}+\chi^{-+}_{0})+\lambda_x\lambda_y\chi^{+-}_{0}\chi^{-+}_{0}}
.
 \nonumber\\ 
%  \label{chiRPA}
\eea
where the interaction parameters are defined  as $\lambda_l=m_l^2 J^{l}_{{\bf q}}$ (here $l=x=y=\parallel, z=\perp$ for the original axes, and $l=z=y=\parallel, x=\perp$ for the rotated axes). 
If the resonance condition should be satisfied for both transverse  parts in Eq.(\ref{chiRPA}) then one must have $m_\parallel^2J_{\bf q}^\perp \approx m_\perp^2J_{\bf q}^\parallel$.
Since the CEF  states and hence the anisotropy  may change as an effect of substitutions in the pure 115 compounds the  resonance signatures may also change 
accordingly. As a result of the sign change of the  superconducting gap function  ($\Delta_{{\bf k}+{\bf Q}}=-\Delta_{{\bf k}}$) at the antiferromagnetic momentum  ${\bf Q}$ the spectral function Im$\chi_0(\bQ,\omega)$ remains  zero for the low frequencies and then shows  a discontinuous jump at the onset frequency of the particle-hole (p-h) continuum, i.e., close to $\omega_c={\rm min}(\Delta_{{\bf k}+{\bf Q}}+\Delta_{{\bf k}})$. This is around $\Delta_0\simeq 2\Delta_1$ , where $2\Delta_1$ is the gap in the superconducting DOS in (Fig.\ref{fig1}.b) which is observed in the tunneling spectrum of \Ce\cite{Rourke:2005}.
 The resonance may appear for energies  $\omega<\omega_c$,  under the condition that   (i) $J^{ll'}_{{\bf q}}{\rm Re} \chi_{0{\bf q}}^{ll'}(\omega)=1$ and (ii)  ${\rm Im} \chi_{0{\bf q}}^{ll'}(\omega)\simeq 0$ ($ll' = +-, -+, zz$).\\

 We  begin our discussion of numerical results by  considering the strongly anisotropic case with  $J^{\bot}_{{\bf Q}}\ll J^{\parallel}_{{\bf Q}}$. Only  in this case, the resonance condition is satisfied  for both $\chi_{0{\bf Q}}^{+-}$ and $\chi_{0{\bf Q}}^{-+}$ components and not for $\chi_{0{\bf Q}}^{zz}$, i.e., a resonance doublet is possible as observed in experiment.
At zero magnetic field a single sharp peak for the degenerate resonance is observed which is shown in Fig.\ref{fig4}. 
By applying the magnetic field the bare $\chi_0^{+-}, \chi_0^{-+}$ susceptibilities  start to split into an upper and lower branch (Figs. \ref{fig2}.b,c), and as a result the two resonance peaks of the doublet are revealed in the RPA frequency spectrum. 
Because   of the anisotropic interaction,  these peaks are completely  distinguishable  for in-plane magnetic field showing a linear Zeeman splitting for small fields. On the other hand they merge together for the out-of-plane field and a single broadened peak appears whose width increases with field. Using the  procedure described above for various fields we obtain  the peak positions of the RPA spectrum versus magnetic field strength in Fig.\ref{fig5}.
  For  field oriented along the ${\hat z}$-direction (Fig.\ref{fig5}.a) we always have a single peak with larger broadening,
and by increasing the magnetic field the peak moves to lower energies.
 But for the magnetic  field applied in ${\hat x}$-direction  (Fig.\ref{fig5}.b) 
 the RPA result shows always two peaks with narrower  line widths. When the field is increased these peaks show first a linear Zeeman splitting and one of them moves to larger and the other one to lower energies. Finally at large field nonlinear behaviour sets in and  both start to move to lower energies.
 These results  are in complete in qualitative agreement with the experimental observation  for \Ce\cite{Stock:12} at lower fields which have sofar been used only.
 It is clear that the linear splitting observed must be modified when larger fields closer to the upper critical fields are applied. The absolute field scale at the maximum $h_B^f=0.7\Delta_0$ is  $B=39.7/(g_f^l\tilde{\chi})$ in Tesla for direction $l =\parallel, \perp$. From a comparison with the experimental linear splitting region up to $6\mbox{T}\ll  H_{c2}^\parallel$ and the one in Fig.\ref{fig5}a with $h_B^f\approx 0.35$ one gets the parameter $g_f^\parallel\tilde{\chi}=3.3$. We note that the reference scale $\Delta_0$ is only constant for fields small to the upper critical field  $( H_{c2}^\parallel=11.9$ T, $H_{c2}^\perp = 4.95$ T). For larger fields one has to scale with $\Delta_0(B)$ which vanishes at the upper critical fields $H_{c2}^{\parallel,\perp}$ where the resonance energies $\omega_r(B)$ also have to vanish for both field directions. When the curves in Fig.~\ref{fig5}a,b are multiplied by the scaling function $\Delta_0(B)$ the field dependence $\omega_r(B)$ is obtained in absolute (meV) units.

 Finally we  mention here that  our results within the phenomenological RPA theory  depend  on the model parameters, i.e. anisotropic matrix elements and quasiparticle interaction energies. Changing these parameters leads to  other interesting regimes, where  the anisotropies of magnetic moment and quasiparticle interaction along the tetragonal axes play the decisive role. The RPA treatment shows that even in absence of the magnetic field an additional peak from the $\chi_{0{\bf q}}^{zz}$ component may appear for suitable parameters and therefore in principle two peaks may exist in the RPA spectrum even at zero field.
 In the  presence of the magnetic field like previous cases  the degenerate (doublet) peak of the  $\chi_{0{\bf q}}^{+-}$ and $\chi_{0{\bf q}}^{-+}$ components splits and finally leading to three different peaks at finite field.
 We should stress here that  by moving to lower interaction energy the resonance  occurs just for $\chi_{0{\bf q}}^{-+}$ and the splitting of the resonance vanishes. This could give an insight into the challenging alternative experiments where  the splitting behavior for in-plane field has not been seen\cite{Panarin:11a}.
 
%%%%%%%%%%%%%%%%%%%%%%%%%%%%%%%%%%%%%%%%%%%%%%%%%%%%%%%%%%%%%%%%%%%%%%%%%%
%%%%%%%%%%%%%%%%%%%%%%%%%%     Section IV     %%%%%%%%%%%%%%%%%%%%%%%%%%%%
%%%%%%%%%%%%%%%%%%%%%%%%%%%%%%%%%%%%%%%%%%%%%%%%%%%%%%%%%%%%%%%%%%%%%%%%%%

 \section{Conclusion}
 \label{secIV}
 We have given an explanation of the observed field splitting of feedback spin resonance excitations recently observed for the first time in the unconventional superconductor \Ce. Our calculations are based on a RPA model with tetragonal anisotropy for collective spin excitations  that may also be relevant for other members of the 115 family.
Using the appropriate hybridized bands and associated Fermi surface as well as the proper $d_{x^2-y^2}$  superconducting gap symmetry the spin resonance appears at $\omega/2\Delta_1= 0.73$, close to the experimental value. Here $2\Delta_1 $ is the main tunneling gap indicated in the DOS of Fig.\ref{fig1}.b and found in Ref.~\onlinecite{Rourke:2005}.

In the isotropic case the resonance is a spin triplet excitation that should split into three modes in a field as predicted in the case of a cuprate model \cite{Ismer:07}. However, because of the presence of strong CEF and hybridization induced anisotropies of g-factors and interactions of 4f-type quasiparticles in 115 compounds the resonance condition may in this case not be fulfilled for all three triplet components. For sufficiently strong anisotropy the zero field resonance is only twofold degenerate transverse doublet state because  the longitudinal component does not satisfy the resonance condition . For field in the tetragonal plane the doublet splits with a linear Zeeman effect for low fields, one branch moving to higher the other two lower energies. However, for larger fields a crossover to nonlinear field dependence with both split resonance energies decreasing sets in. For field perpendicular to the tetragonal plane no splitting but only a broadening of the doublet resonance excitation appears. These salient features of our model calculation correspond closely to the experimental observations in \Ce~  performed for small fields \cite{Stock:12}. It would be very interesting to investigate the predicted nonlinear field dependence for larger fields.

Finally we note that other scenarios are possible depending on the anisotropies and strengths of quasiparticle interactions where for example the in-plane splitting of the resonance disappears and a single peak with anomalous broadening as for the out-of-plane field is observed. In fact this behaviour was proposed in an alternative experiment \cite{Panarin:11a} and further experimental as well as theoretical investigations are necessary to clarify this issue.

  %%%%%%%%%%%%%%%%%%%%%%%%%%%%%%%%%%%%%%%%%%%%%%%%%%%%%%%%%%%%%%%%%%%%%%%%%%
%%%%%%%%%%%%%%%%%%%%%%%%%%     Section IV     %%%%%%%%%%%%%%%%%%%%%%%%%%%%
%%%%%%%%%%%%%%%%%%%%%%%%%%%%%%%%%%%%%%%%%%%%%%%%%%%%%%%%%%%%%%%%%%%%%%%%%%
% \bibliographystyle{prl}
\bibliography{CeCoIn5}

%%%%%%%%%%%%%%%%%%%%%%%%%%%%%%%%%%%%%%%%%%%%%%%%%%%%%%%%%%%%%%%%%%%%%%%%%%
%%%%%%%%%%%%%%%%%%%%%%%%%%%%%      References        %%%%%%%%%%%%%%%%%%%%%

\end{document}